\documentclass[sigconf]{acmart}

\newcommand{\dquotes}[1]{``#1''}

\usepackage{booktabs}
\usepackage{hyperref}
 \usepackage{array} 
\usepackage{multirow}
\usepackage{colortbl} 
\usepackage{hhline} 
\usepackage{makecell} 
\usepackage{tabularx}
\usepackage{arydshln}
\usepackage{longtable} 
\usepackage{array} 
\usepackage{hyperref} 

\usepackage[most]{tcolorbox} 
\usepackage{enumitem}

\definecolor{myviolet}{rgb}{0.9,0.1,0.6}

\definecolor{bluegreen}{rgb}{0,0.5,0.5}

\copyrightyear{2024}
\acmYear{2024}
\setcopyright{rightsretained}
\acmConference[KDD '24]{Proceedings of the 30th ACM SIGKDD Conference on Knowledge Discovery and Data Mining}{August 25--29, 2024}{Barcelona, Spain}
\acmBooktitle{Proceedings of the 30th ACM SIGKDD Conference on Knowledge Discovery and Data Mining (KDD '24), August 25--29, 2024, Barcelona, Spain}\acmDOI{10.1145/3637528.3671474}
\acmISBN{979-8-4007-0490-1/24/08}

\makeatletter
\gdef\@copyrightpermission{
  \begin{minipage}{0.3\columnwidth}
   \href{https://creativecommons.org/licenses/by/4.0/}{\includegraphics[width=0.90\textwidth]{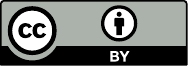}}
  \end{minipage}\hfill
  \begin{minipage}{0.7\columnwidth}
   \href{https://creativecommons.org/licenses/by/4.0/}{This work is licensed under a Creative Commons Attribution International 4.0 License.}
  \end{minipage}
  \vspace{5pt}
}
\makeatother

\begin{document}

\title{A Review of Modern Recommender Systems Using Generative Models (Gen-RecSys)}

\author{Yashar Deldjoo}
\affiliation{%
  \institution{Polytechnic University of Bari}
  \city{Bari}
  \country{Italy}}
\email{deldjooy@acm.org}

\author{Zhankui He}
\affiliation{%
  \institution{University of California}
  \city{La Jolla}
  \country{USA}}
\email{zhh004@ucsd.edu}

\author{Julian McAuley}
\affiliation{%
  \institution{University of California}
  \city{La Jolla}
  \country{USA}}
\email{jmcauley@ucsd.edu}

\author{Anton Korikov}
\affiliation{%
  \institution{University of Toronto}
  \city{Toronto}
  \country{Canada}}
\email{anton.korikov@mie.utoronto.ca}

\author{Scott Sanner}
\affiliation{%
  \institution{University of Toronto}
  \city{Toronto}
  \country{Canada}}
\email{ssanner@mie.utoronto.ca}

\author{Arnau Ramisa}
\affiliation{%
  \institution{Amazon*\thanks{*This work does not relate to the author's position at Amazon.}}
  \city{Palo Alto}
  \country{USA}}
\email{aramisay@amazon.com}

\author{Ren\'e Vidal}
\affiliation{%
  \institution{Amazon*}
  \city{Palo Alto}
  \country{USA}}
\email{vidalr@seas.upenn.edu}

\author{Maheswaran Sathiamoorthy}
\affiliation{%
  \institution{Bespoke Labs}
  \city{Santa Clara}
  \country{USA}}
\email{mahesh@bespokelabs.ai}

\author{Atoosa Kasirzadeh}
\affiliation{%
  \institution{University of Edinburgh}
  \city{Edinburgh}
  \country{UK}}
\email{atoosa.kasirzadeh@gmail.com}

\author{Silvia Milano}
\affiliation{%
  \institution{University of Exeter and LMU Munich}
  \city{Munich}
  \country{Germany}}
\email{milano.silvia@gmail.com}

\renewcommand{\shortauthors}{Yashar Deldjoo et al.}

\begin{abstract} 
Traditional recommender systems typically use user-item rating histories as their main data source. However, deep generative models now have the capability to model and sample from complex data distributions, including user-item interactions, text, images, and videos, enabling novel recommendation tasks. This comprehensive, multidisciplinary survey connects key advancements in RS using Generative Models (Gen-RecSys), covering: interaction-driven generative models; the use of large language models (LLM) and textual data for natural language recommendation; and the integration of multimodal models for generating and processing images/videos in RS. Our work highlights necessary paradigms for evaluating the impact and harm of Gen-RecSys and identifies open challenges. This survey accompanies a \textbf{tutorial} presented at ACM KDD'24, with supporting materials provided at: \href{https://encr.pw/vDhLq}{https://encr.pw/vDhLq}.

\end{abstract}



\maketitle

\section{Introduction}
Advancements in generative models have significantly impacted the evolution of recommender systems (RS). Traditional RS, which relied on capturing user preferences and item features within a specific domain — often referred to as \dquotes{\textit{narrow experts}} -- are now being complemented and, in some instances, surpassed by generative models. These models have introduced innovative ways of conceptualizing and implementing recommendations. Specifically, modern generative models learn to represent and sample from complex data distributions, including not only user-item interaction histories but also text and image content, unlocking these data modalities for novel and interactive recommendation tasks. 

\begin{figure*}[!t]
    \centering
    \includegraphics[width = 0.90\linewidth]{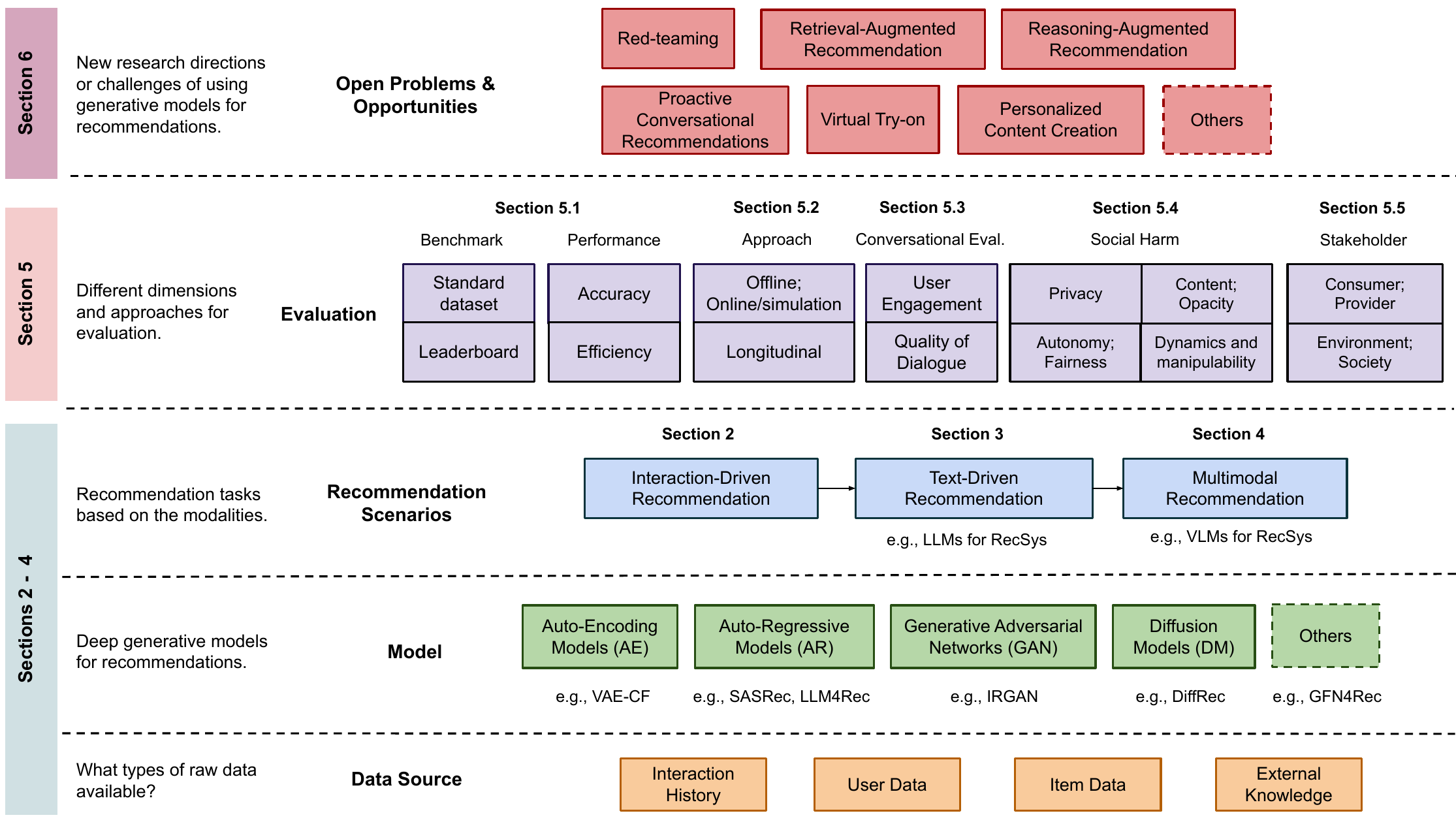}
    \caption{
    Overview of the areas of interest in generative models in recommendation.
    }
    \label{fig:chart}
\end{figure*}

Moreover, advances in natural language processing (NLP) through the introduction of large language models (LLMs) such as ChatGPT~\cite{chatgpt} and Gemini~\cite{team2023gemini} have showcased remarkable \textit{emergent} capabilities~\cite{wei2022emergent}, including reasoning, in-context few-shot learning, and access to extensive open-world information within their pre-trained parameters. Because of their broad generalist abilities, these pretrained generative models have opened up an exciting new research space for a wide variety of recommendation applications (see Table~\ref{tab:applications}), e.g.,  enhanced personalization, improved conversational interfaces, and richer explanation generation, among others.

The core of generative models lies in their ability to model and sample from their training \textit{data distribution} for various inferential purposes, which enables two primary modes of application for RS:

\begin{enumerate}[leftmargin=8pt]

\item \textbf{Directly trained models.} This approach trains generative models, such as VAE-CF (Variational AutoEncoders for Collaborative Filtering)~\cite{liang2018variational} (cf. Section~\ref{subsec:ui_ae}) directly on user-item interaction data to predict user preferences, without using large, diverse pretraining datasets. These models learn the probability distribution of items a user might like based on their previous interactions.


\item \textbf{Pretrained models.} 
This strategy uses models pretrained on diverse data (text, images, videos) to understand complex patterns, relationships, and contexts that often exhibit (emergent) generalization abilities to a range of novel tasks~\cite{wei2022emergent}. Among a variety of applications, this survey covers the use of pretrained Gen-RecSys models in the following settings:
\begin{itemize}
\item \textit{Zero- and Few-shot Learning} (cf. Section~\ref{subsec:ICL}), using in-context learning (ICL) for broad understanding without extra training.
\item \textit{Fine-Tuning} (cf. Section \ref{sec:RARec}), adjusting model parameters using specific datasets for tailored recommendations. 
\item \textit{Retrieval-Augmented Generation (RAG)} (cf. Section \ref{sec:RARec}), integrating information retrieval with generative modeling for contextually relevant outputs. 
\item \textit{Feature Extraction for Downstream Recommendation} (cf. Section \ref{sec:LLM 4 RS Inputs}), e.g., generating embeddings or token sequences for complex content representation.
\item \textit{Multimodal Approaches} (cf. Section \ref{sec:MM-LMMs}), jointly using multiple data types such as text, image, and video to enhance and improve the recommendation experience. 
\end{itemize}
\end{enumerate}

\begin{table*}[!t]
\centering
\caption{Example applications of Gen-RecSys methods.}
\label{tab:applications}
\renewcommand{\arraystretch}{1.24} 
\begin{tabular}{p{0.7\textwidth}p{0.2\textwidth}}
\toprule
\textbf{Description} & \textbf{Relevant Section} \\
\midrule
Utilize text in RS, including item descriptions, user preferences, reviews, queries, and conversation histories. Examples include generative and conversational recommendations and explanations. & Sections \ref{sec:LLM gen rec}, \ref{sec:ConvRec} \\ 

Using and generating images for recommendation, reasoning, and content creation. & Section 4, Sections \ref{sec:gener-mm} \\ 

Applying the emergent reasoning abilities of pre-trained models to recommendation tasks, including in-context learning and tool-augmented reasoning. & Sections \ref{sec:RARec}, \ref{sec:ConvRec} \\ 

Integrating RS with external knowledge sources through retrieval augmented generation. & Section \ref{sec:RARec} \\

Selecting or generating informative user-item interactions to improve RS model training. & Sections \ref{subsec:ui_gan}, \ref{subsec:ui_dm} \\ 
Generating recommendation results with complex structures such as list-wise or page-wise outputs. & Sections \ref{subsec:ui_ae}, \ref{subsec:ui_ar}, \ref{subsec:ui_gan}, \ref{subsec:ui_other} \\ 

Facilitating conversational recommendation through full NL dialogue. & Section \ref{sec:ConvRec} \\
\bottomrule
\end{tabular}
\renewcommand{\arraystretch}{1} 
\end{table*}

\subsection{Recent Surveys and Our Contributions}
\subsubsection*{Recent Relevant Surveys}
Recent surveys have marked significant advancements in the field. We highlight our contributions and distinguish our survey by its comprehensive and unique approach.

\begin{itemize}[leftmargin=*]
\item \citet{deldjoo2021survey} explore GAN-based RS across four different recommendation scenarios (graph-based, collaborative, hybrid, context-aware).
    \item \citet{li2023large} explore training strategies and learning objectives of LLMs for RS. 
    \item \citet{wu2023survey} discuss both the use of LLMs to generate RS input tokens or embeddings as well as the use of LLMs \textit{as} an RS;
    \item \citet{lin2023can} focus on adapting LLMs in RS, detailing various tasks and applications. \citet{fan2023recommender} overview LLMs in RS, emphasizing pre-training, fine-tuning, and prompting, while \citet{vats2024exploring} review LLM-based RS, introducing a heuristic taxonomy for categorization.
    \item \citet{huang2024foundation}, explore using foundation models (FMs) in RS.
\item \citet{wang2023generative} introduce GeneRec, a next-gen RS that personalizes content through AI generators and interprets user instructions to gather user preferences.
\end{itemize}
While the mentioned surveys offer crucial insights, their scope is often limited to LLMs~\cite{li2023large,wu2023survey,lin2023can,fan2023recommender,vats2024exploring} or, more broadly, FMs~\cite{huang2024foundation} and/or specific models such as GANs~\cite{deldjoo2021survey}, without considering the wider spectrum of generative models and data modalities. The work by~\cite{wang2023generative}  provides a more relevant survey on Gen-RecSys although their work is mostly on 
personalized content generation.

\subsubsection*{Core Contributions.}

Figure \ref{fig:chart} illustrates the structure of our Gen-RecSys survey. It categorizes data sources, recommendation models, and scenarios, extending to system evaluation and challenges. We present a \textbf{systematic} approach to deconstructing the Gen-RecSys recommendation process into distinct components and methodologies. Our contributions are summarized as follows.
\begin{enumerate}
    \item Our survey is broader in scope than the surveys mentioned above, encompassing not just LLMs but a wide array of generative models in RS.
    \item We have chosen to classify these models based on the type of data and modality they are used for, such as user-item data (cf. Section \ref{sec:UI-GEN}), text-driven (cf. Section \ref{subsec:RecLLMs}), and multimodal (cf. Section \ref{sec:MM-LMMs}) models, as shown in the \emph{Rec. Scenario} layer. 
    \item Within each modality discussion, we provide an in-depth exploration of deep generative model paradigms as shown in the \emph{Model} layer, yet with a broader scope that spans multiple contexts and use cases, offering a critical analysis of their roles and effectiveness in respective sections. 
    \item We study the evaluation of Gen-RecSys with finer details, shedding light on multiple aspects such as benchmarks, evaluation for impact and harm relative to multiple stakeholders, and conversational evaluation. This evaluation framework is particularly notable as it helps to understand the complex challenges intrinsic to Gen-RecSys.
    \item We discuss several open research challenges and issues. Our survey benefits from the expertise of scholars/industry practitioners from diverse institutions and disciplines.
\end{enumerate}

\section{Generative Models for Interaction-\\Driven Recommendation}
\label{sec:UI-GEN}


Interaction-driven recommendation is a setup where only the user-item interactions (e.g.,~``user A clicks item B'') are available, which is the most general setup studied in RS. In this setup, we concentrate on the inputs of user-item interactions and outputs of item-recommended lists or grids rather than richer inputs or outputs from other modalities such as textual reviews. Even though no textual or visual information is involved, generative models~\cite{goodfellow2014generative, kingma2013auto, hochreiter1997long, vaswani2017attention, sohl2015deep} 
still show their unique usefulness.
In this section, we examine the paradigms of generative models for recommendation tasks with user-item interactions, including auto-encoding models~\cite{kingma2013auto}, auto-regressive models~\cite{vaswani2017attention, hochreiter1997long}, generative adversarial networks~\cite{goodfellow2014generative}, diffusion models~\cite{sohl2015deep} and more.

\subsection{Auto-Encoding Models} \label{subsec:ui_ae}

Auto-encoding models learn to reconstruct their inputs. This capability allows them to be used for various purposes, including denoising, representation learning, and generation tasks.

\subsubsection{Preliminaries: Denoising Auto-Encoding Models} \label{sec:denoising ae}

Denoising Auto-Encoding models are a group of models that learn to recover the original inputs from a corrupted version of the inputs.
Traditionally, denoising auto-encoding models refer to a group of Denoising Autoencoders~\cite{vincent2008extracting, sedhain2015autorec} with hidden layers as a ``bottleneck''. For example, AutoRec~\cite{sedhain2015autorec} tries to reconstruct the input vector, which is partially observed.
More broadly, BERT-like models~\cite{devlin2018bert,sun2019bert4rec,wu2016collaborative} are also treated as denoising auto-encoding models. Such models recover corrupted (i.e., masked) inputs through stacked self-attention blocks~\cite{sun2019bert4rec,he2021locker}. For example, BERT4Rec~\cite{sun2019bert4rec} is trained to predict masked items in given user historical interaction sequences. Therefore, BERT-like~\cite{devlin2018bert} models can be used for next-item prediction in the inference phase~\cite{sun2019bert4rec,he2021locker}.

\subsubsection{Variational Auto-Encoding Models}
\label{sec:vae}

Variational Autoencoders (VAEs) are models that learn stochastic mappings from an 
input $x$ from a often complicated probability distribution $p$ to a
probability distribution $q$. This distribution, $q$, is typically simple (e.g., a normal distribution), enabling the use of a decoder to generate outputs $\hat{x}$ by sampling from $q$~\cite{kingma2013auto}.  VAEs find wide applications in traditional RS, particularly for collaborative filtering~\cite{liang2018variational}, sequential recommendation~\cite{sachdeva2019sequential} and slate generation~\cite{jiang2018beyond,liu2021variation,deffayet2023generative}. Compared to Denoising Autoencoders, VAEs often demonstrate superior 
performance in collaborative filtering due to stronger modeling assumptions, 
such as VAE-CF~\cite{liang2018variational}. Additionally, Conditional VAE (CVAE)~\cite{sohn2015learning} models learn distributions of preferred recommendation lists for a given user. This makes them useful for generating those lists beyond a greedy ranking schema.  Examples like ListCVAE~\cite{jiang2018beyond} and PivotCVAE~\cite{liu2021variation} use VAEs to generate entire recommendation lists rather than solely ranking individual items.

\subsection{Auto-Regressive Models} \label{subsec:ui_ar}

Given an input sequence $\mathbf{x}$, at step $i$, auto-regressive models~\cite{bengio2000neural} learn the conditional probability distribution 
$p(x_i|\mathbf{x}_{<i})$, where $\mathbf{x}_{<i}$ represents the subsequence before step $i$. Auto-regressive models are primarily used for sequence modeling~\cite{bengio2000neural,vaswani2017attention,van2016wavenet}. In RS, they find wide applications in session-based or sequential recommendations~\cite{hidasi2015session,kang2018self}, model attacking~\cite{yue2021black}, and bundle recommendations~\cite{bai2019personalized, hu2019sets2sets}, with recurrent neural networks~\cite{hidasi2015session, bai2019personalized, hu2019sets2sets}, self-attentive models~\cite{kang2018self}, and more.

\subsubsection{Recurrent Auto-Regressive Models}

Recurrent neural networks (RNNs)~\cite{hochreiter1997long, chung2014empirical} have been use to predict the next item in session-based and sequential recommendations, such as GRU4Rec~\cite{hidasi2015session} and its variants~\cite{hidasi2018recurrent, yue2024linear} (e.g., predicting the next set of items in basket or bundle recommendations, such as set2set~\cite{hu2019sets2sets} and BGN~\cite{bai2019personalized}). Moreover, using the auto-regressive generative nature of recurrent networks, researchers extract model-generated user behavior sequences, which are used in the research of model attacking~\cite{yue2021black}.

\subsubsection{Self-Attentive Auto-Regressive Models}

Self-attentive models replace the recurrent unit with self-attention and related modules, inspired by transformers~\cite{vaswani2017attention}. This group of models can be used in session-based recommendation and sequential recommendation~\cite{kang2018self,wu2020sse,lin2020fissa, petrov2023generative}, next-basket or bundle prediction~\cite{yu2020predicting}, and model attacking~\cite{yue2021black}. Meanwhile, the benefits of self-attentive models are that they handle long-term dependencies better than RNNs and enable parallel training~\cite{vaswani2017attention}. Additionally, self-attentive models are the \emph{de-facto} option for pre-trained models~\cite{devlin2018bert} and large language models~\cite{brown2020language, wei2022emergent,sparks_of_agi}, which is gaining traction in RS. More details about using such language models for recommendations will be discussed in Section~\ref{subsec:RecLLMs}.

\subsection{Generative Adversarial Networks} \label{subsec:ui_gan}

Generative adversarial networks (GANs)~\cite{mirza2014conditional,goodfellow2014generative} are composed of two primary components: a generator network and a discriminator network. These networks engage in adversarial training to enhance the performance of both the generator and the discriminator. GANs are used in RS for multiple purposes~\cite{wang2017irgan,cai2018kbgan,chen2019generative}. In the interaction-driven setup, GANs are proposed for selecting informative training samples~\cite{cai2018kbgan, wang2017irgan}, for example, in IRGAN~\cite{wang2017irgan, wang2018neural}, the
generative retrieval model is leveraged to sample negative items. Meanwhile, GANs synthesize user preferences or interactions to augment training data~\cite{chae2018cfgan, wang2019enhancing}. Additionally, GANs have shown effectiveness in generating recommendation lists or pages, such as~\cite{chen2019generative} in whole-page recommendation settings.

\subsection{Diffusion Models} \label{subsec:ui_dm}
\label{sec:diffusion}

Diffusion models ~\cite{sohl2015deep} generate outputs through a two-step process: (1) corrupting inputs into noise via a forward process, and (2) learning to recover the original inputs from the noise iteratively in a reverse process. Their impressive generative capabilities have attracted growing interest from the RS community.

First, a group of works~\cite{wang2023diffusion, walker2022recommendation} learns users' future interaction probabilities through diffusion models. For example, DiffRec~\cite{wang2023diffusion} predicts users' future interactions using corrupted noises from the users' historical interactions. Second, another group of works~\cite{liu2023diffusion, wu2023diff4rec} focuses on diffusion models for training sequence augmentation, showing promising results in alleviating the data sparsity and long-tail user problems in sequential recommendation.

\subsection{Other Generative Models} \label{subsec:ui_other}

In addition to the previously mentioned generative models, RS also draw upon other types of generative models. For instance, VASER~\cite{zhong2020session} leverages 
normalizing flows~\cite{rezende2015variational} (and 
VAEs~\cite{kingma2013auto}) for session-based recommendation. GFN4Rec~\cite{liu2023generative}, on the other hand, adapts generative flow networks~\cite{bengio2021flow,pan2023better} for listwise recommendation. Furthermore, IDNP~\cite{du2023idnp} utilizes generative neural processes~\cite{garnelo2018neural,garnelo2018conditional} for sequential recommendation. In summary, various generative models are explored in RS, even in settings without textual or visual modalities.

\section{Large Language Models in Recommendation}
\label{subsec:RecLLMs}
While language has been leveraged by content-based RS for over three decades \cite{lops2011content}, the advent of pretrained LLMs and their emergent abilities for generalized, multi-task natural language (NL) reasoning \cite{brown2020language, wei2022emergent,sparks_of_agi} has ushered in a new stage of language-based recommendation. Critically, NL constitutes a unified, expressive, and interpretable medium that can represent not only item features or user preferences, but also user-system interactions, recommendation task descriptions, and external knowledge \cite{geng2022recommendation}. For instance, items are often associated with rich text including titles, descriptions, semi-structured textual metadata, and reviews. Similarly, user preferences can be articulated in NL in many forms, such as reviews, search queries, liked item descriptions, and dialogue utterances. 

Pretrained LLMs provide new ways to exploit this textual data: recent research (e.g., \cite{sanner2023large, sileo2022zero, geng2022recommendation, he2023large, friedman2023leveraging}) has shown that in many domains, LLMs have learned useful reasoning abilities for making and explaining item recommendations based on user preferences as well as facilitating conversational recommendation dialogues. As discussed below, these pretrained abilities can be further augmented through prompting (e.g., \cite{sanner2023large,sileo2022zero, liu2023chatgpt}), fine-tuning (e.g., \cite{geng2022recommendation, harte2023leveraging, kang2023llms,zhang2023collm}), retrieval (e.g., \cite{hou2023large, wang2023zero,dai2023uncovering,friedman2023leveraging, kemper2024retrieval}), and other external tools (e.g., \cite{wang2022unicrs, friedman2023leveraging, zeng2024automated}. 

We next proceed to survey the developments in LLM-based RS's, first discussing encoder-only LLMs for dense retrieval and cross-encoding (Section \ref{sec:encoder-only LLM}) followed by generative NL recommendation and explanation with sequence-to-sequence (seq2seq) LLMs (Section \ref{sec:LLM gen rec}). We then review the complementary use of RS and LLMs covering RAG (Section \ref{sec:RARec}) and LLM-based feature extraction (Section \ref{sec:LLM 4 RS Inputs}), before concluding with a review of conversational recommendation methods (Section \ref{sec:ConvRec}).


\subsection{Encoder-only LLM Recommendation} \label{sec:encoder-only LLM}
\subsubsection{Recommendation as Dense Retrieval} \label{sec:dense retrieval}

A common task is to retrieve the most relevant items given a NL preference statement using item texts, for which dense retrieval has become a key tool. Dense retrievers \citep{fan2022pre}
produce a ranked list of documents given a query by evaluating the similarity (e.g., dot product or cosine similarity) between encoder-only LLM document embeddings and the query embedding. They are highly scalable tools
(especially when used with approximate search libraries like
FAISS\footnote{\href{https://github.com/facebookresearch/faiss}{https://github.com/facebookresearch/faiss}})
because documents and queries are encoded separately, allowing for dense vector 
indexing of documents before querying. To use dense retrieval for recommendation \citep{penha2020does}, first, a component of each item's text content, such as its title, description, reviews, etc., is treated as a document and a dense item index is constructed. Then, a query is formed by some NL user preference description, for instance: an actual search query, the user's recently liked item titles, or a user utterance in a dialogue. 

Several recent works explore recommendation \textit{as} standard dense retrieval with retrievers that are off-the-shelf \citep{penha2020does, harte2023leveraging, zhang2023recipe} and fine-tuned \citep{mysore2023large,li2023gpt4rec,hou2023large}. More complex dense retrieval methods include review-based retrieval with contrastive BERT fine-tuning \citep{abdollah2023self} and multi-aspect query decomposition \cite{korikov2024multi}, and the use of a second-level encoder to fuse the embedding of a user's recently liked items into a user embedding before scoring \citep{wu2019neural, li2022miner}.


\subsubsection{Recommendation via LLM Item-Preference Fusion} \label{sec:Item-Preference Fusion}
Several works approach rating prediction by \textit{jointly} embedding NL item and preference descriptions in LLM cross-encoder architectures with an MLP rating prediction head \cite{zhang2021unbert,yao2022reprbert, wu2021empowering,qiu2021u, zhang2023prompt}. Such fusion-in-encoder methods often exhibit strong performance because they allow interaction between user and item representations, but are much more computationally expensive than dense retrieval
and thus may be best used for small item sets or as rerankers  \cite{mysore2023large}. 

%

\subsection{LLM-based Generative Recommendation}
 
\label{sec:LLM gen rec}

In LLM-based generative recommendation, tasks are expressed as token sequences -- called \emph{prompts} -- which form an input to a seq2seq LLM. The LLM then generates another token sequence to address the task -- with example outputs including: a recommended list of item titles/ids
\cite{mao2023unitrec, harte2023leveraging, sanner2023large, sileo2022zero}, a rating \cite{bao2023tallrec,kang2023llms}, or an explanation \cite{ni2019justifying, li2020generate, hada2021rexplug, geng2022recommendation, li2023personalized}.
These methods rely on the pretraining of LLMs on large text corpora to provide knowledge about a wide range of entities, human preferences, and commonsense reasoning that can be used directly for recommendation or leveraged to improve generalization and reduce domain-specific data requirements for fine-tuning or prompting \cite{wei2022emergent, sparks_of_agi}.


\subsubsection{Zero- and Few- Shot Generative Recommendation}\label{subsec:ICL}
Several recent publications \cite{kang2023llms, sanner2023large,sileo2022zero, liu2023chatgpt} have evaluated with off-the-shelf LLM generative recommendation, focusing on domains that are prevalent in the LLM pre-training corpus such as movie and book recommendation. Specifically, these methods construct a prompt with a NL description of user preference (often using a sequence of recently liked item titles) and an instruction to recommend the next $k$ item titles \cite{sanner2023large,sileo2022zero,liu2023chatgpt} or predict a rating \cite{kang2023llms, liu2023chatgpt}. While, overall, untuned LLMs underperform supervised CF methods trained on sufficient data \cite{kang2023llms,sileo2022zero}, they are competitive in near cold-start settings \cite{sanner2023large, sileo2022zero}. Few-shot prompting (or in-context learning), in which a prompt contains examples of input-output pairs, typically outperforms zero-shot prompting \cite{sanner2023large}. 

\subsubsection{Tuning LLMs for Generative Recommendation} To improve an LLM's generative recommendation performance and add knowledge to its internal parameters,
multiple works focus on fine-tuning \cite{geng2022recommendation, bao2023tallrec, harte2023leveraging, mao2023unitrec, kang2023llms} and prompt-tuning \cite{li2023personalized, cui2022m6,zhang2023collm} strategies. Recent works fine-tune LLMs 
on NL input/output examples constructed from user-system interaction history and task descriptions for rating prediction \cite{bao2023tallrec,kang2023llms} and sequential recommendation \cite{mao2023unitrec, harte2023leveraging}, or in the case of P5 \cite{geng2022recommendation}, both preceding tasks plus top-$k$ recommendation, explanation generation, and review summarization. Other recommendation works study prompt tuning approaches \cite{li2023personalized, cui2022m6, zhang2023collm}, which adjust LLM behaviour by tuning a set of continuous (or soft) prompt vectors as an alternative to tuning internal LLM weights.
 
\paragraph{Generative Explanation} A line of recent work focuses on explanation generation where training explanations are extracted from reviews, since reviews often express reasons why a user decided to interact with an item. Techniques include fine-tuning \citep{geng2022recommendation, li2023personalized, wang2024deciphering}, prompt-tuning \citep{li2023personalized, li2020generate}, chain-of-thought prompting \citep{rahdari2024logic}, and controllable decoding \citep{ni2018personalized,ni2019justifying,hada2021rexplug,xie2023factual} -- where additional predicted parameters such as ratings steer LLM decoding.


\subsection{Retrieval Augmented Recommendation} \label{sec:RARec}
Adding knowledge to an LLM internal memory through tuning can improve performance, but it requires many parameters and re-tuning for every system update. An alternative is retrieval-augmented generation (RAG) \cite{lewis2020retrieval, izacard2020leveraging, borgeaud2022improving}, which conditions output on information from an external source such as a dense retriever (Section \ref{sec:encoder-only LLM}). RAG methods facilitate online updates, reduce hallucinations, and generally require fewer LLM parameters since knowledge is externalized \cite{mialon2023augmented,borgeaud2022improving, izacard2020leveraging}.



 RAG has recently begun to be explored for recommendation, with the most common approach being to first use a retriever or RS to construct a candidate item set based on a user query or interaction history, and then prompt an encoder-decoder LLM to rerank the candidate set \citep{hou2023large, wei2024llmrec, wang2023zero,dai2023uncovering, yang2022improving}. For RAG-based explanation generation, \citeauthor{xie2023factual} \cite{xie2023factual} generate queries based on interaction history to retrieve item reviews which are used as context to generate an explanation of the recommendation. RAG is also emerging as a key paradigm in conversational recommendation (c.f. Sec \ref{sec:ConvRec}): for example, RAG is used in \cite{friedman2023leveraging} to retrieve relevant user preference descriptions from a user ``memory'' module to guide dialogue, and by \citeauthor{kemper2024retrieval} \cite{kemper2024retrieval} to retrieve information from an item's reviews to answer user questions. 

\subsection{LLM-based Feature Extraction} \label{sec:LLM 4 RS Inputs} Conversely to how RS or retrievers are used in RAG to obtain inputs for LLMs (Section \ref{sec:RARec}), LLMs can also be used to generate inputs for RS \cite{harte2023leveraging, he2022query, rajput2024recommender, yuan2023go, mysore2023large, li2023gpt4rec}. For instance: LLM2-BERT4Rec \cite{harte2023leveraging} initializes BERT4Rec (Section \ref{sec:denoising ae}) item embeddings of item texts;
Query-SeqRec \cite{he2022query} includes LLM query embeddings as inputs to a transformer-based recommender; and TIGER \cite{rajput2024recommender} first uses an LLM to embed item text, then quantizes this embedding into a semantic ID, and finally trains a T5-based RS to generate new IDs given a user's item ID history. Similarly, MINT \cite{mysore2023large} and GPT4Rec \cite{li2023gpt4rec} produce inputs for a dense retriever by prompting an LLM to generate a query given a user's interaction history.

\subsection{Conversational Recommendation} \label{sec:ConvRec}
The recent advances in LLMs have made fully NL system-user dialogues a feasible and novel recommendation interface, bringing in a new stage of conversational recommendation (ConvRec) research. This direction studies the application of LLMs in multi-turn, multi-task, and mixed-initiative NL recommendation conversations \cite{friedman2023leveraging, jannach2020survey}, introducing dialogue history as a rich new form of interaction data. Specifically, ConvRec includes the study and integration of diverse conversational elements such as dialogue management, recommendation, explanation, QA, critiquing, and preference elicitation \cite{jannach2020survey, sanner:www21}. While some research \cite{he2023large} approaches ConvRec with a monolithic LLM such as GPT4, other works rely on an LLM to facilitate NL dialogue \textit{and} integrate calls to a recommender module which generates item recommendations based on dialogue or interaction history \cite{li2018conversational, kang2019recommendation, chen2019towards, yang2022improving, wang2022unicrs, austin2024bayesian, handa2024bayesian}. Further research advances ConvRec system architectures with multiple tool-augmented LLM modules, incorporating components for dialogue management, explanation generation, and retrieval \cite{friedman2023leveraging, gao2023chat,wang2023recmind, kemper2024retrieval, zeng2024automated, joko2024doing}.



%


\section{Generative Multimodal Recommendation Systems}
\label{sec:MM-LMMs}
In recent years, users have come to expect richer interactions than simple text or image queries. For instance, they might provide a picture of a desired product along with a natural language modification (e.g., a dress like the one in the picture but in red). Additionally, users want to visualize recommendations to see how a product fits their use case, such as how a garment might look on them or how a piece of furniture might look in their room. These interactions require new RS that can discover unique attributes in each modality.  In this section, we discuss RS that utilize multiple data modalities. In Sections~\ref{sec:motivations-mm}-\ref{sec:challenges-mm} we discuss motivations and challenges to the design of multimodal RS. In Sections~\ref{sec:discrim-mm}-\ref{sec:gener-mm} we review contrastive and generative approaches to multimodal RS, respectively. 


\subsection{Why Multimodal Recommendation?}
\label{sec:motivations-mm}

Retailers often have multimodal information about their customers and products, including product descriptions, images and videos, customer reviews and purchase history. However, existing RS typically process each source independently and then combine the results by fusing unimodal relevance scores. 

In practice, there are many use cases in which such a ``late fusion'' approach may be insufficient to satisfy the customer needs. One such use case is the \emph{cold start problem}: when user behavioral data cannot be used to recommend existing products to new customers, or new products to existing customers, it is useful to gather diverse information about the items so that preference information can be transferred from existing products or customers to new ones.

Another use case occurs when different modalities are needed to understand the user request. For example, to answer the request “best metal and glass black coffee table under \$300 for my living room”, the system would need to reason about the appearance and shape of the item in context with the appearance and shape of other objects in the customer room, which cannot be achieved by searching with either text or image independently. Other examples of multimodal requests include an image or audio of the desired item together with text modification instructions (e.g., a song like the sound clip provided but in acoustic), or a complementary related product (e.g., a kickstand for the bicycle in the picture). 

A third use case for multimodal understanding is in RS with complex outputs, such as virtual try-on features or intelligent multimodal conversational shopping assistants.


\subsection{Challenges to Multimodal Recommendation}
\label{sec:challenges-mm}

The development of multimodal RS faces several challenges. 
First, collecting data to train multimodal systems (e.g., image-text-image triplets) is significantly harder than for unimodal systems. As a result, annotations for some modalities may be incomplete~\cite{rahate2022multimodal}. 

Second, combining different data modalities to improve recommendation results is not simple. For instance, existing contrastive learning approaches~\cite{radford2021learning, jia2021scaling, li2021align, li2022blip} map each data modality to a common latent space in which all modalities are approximately aligned. However, such approaches often capture information that is shared across modalities (e.g., text describing visual attributes), but they overlook complementary aspects that could benefit recommendations (e.g., text describing non visual attributes)~\cite{guo2019deep}. In general we would like the modalities to compensate for one another and result in a more complete joint representation. While fusion-based approaches~\cite{li2021align, li2022blip} do learn a joint multimodal representation, ensuring the alignment of information that is shared and leaving some flexibility to capture complementary information across modalities remains a challenge. Third, learning multimodal models requires orders of magnitude more data than learning models for individual data modalities.

Despite these challenges, we believe multimodal generative models will become the standard approach. Indeed, recent literature shows significant advances on the necessary components to achieve effective multimodal generative models for RS, including
(1) the use of LLMs and diffusion models to generate synthetic data for labeling purposes~\cite{brooks2023instructpix2pix, rosenbaum2022clasp, nguyen2024dataset},
(2) high quality unimodal encoders and decoders~\cite{he2022masked, kirillov2023segment}, (3) better techniques for aligning the latent spaces from multiple modalities into a shared one~\cite{radford2021learning, li2022blip, girdhar2023imagebind}, 
(4) efficient re-parametrizations and training algorithms~\cite{jang2016categorical}, 
and (5) techniques to inject structure to the learned latent space to make the problem tractable~\cite{sohl2015deep}. 


\subsection{Contrastive Multimodal Recommendation}
\label{sec:discrim-mm}


As discussed before~\ref{sec:challenges-mm}, learning multimodal generative models is very difficult because we need to not only learn a latent representation for each modality but also ensure that they are aligned. One way to address this challenge is to first learn an alignment between multiple modalities and then learn a generative model on ``well-aligned'' representations. In this subsection, we discuss two representative contrastive learning approaches: CLIP and ALBEF.

\emph{Contrastive Language-Image Pre-training (CLIP)~\cite{radford2021learning}} is a popular approach, in which the task is to project images and associated text into the same point of the embedding space with parallel image and text encoders. This is achieved with a symmetric cross-entropy loss over the rows and columns of the cosine similarity matrix between all possible pairs of images and text in a training minibatch. 


\emph{Align Before you Fuse (ALBEF)}
\cite{li2021align} augments CLIP with a multimodal encoder that fuses the text and image embeddings, and proposes three objectives to pre-train the model: Image-text contrastive learning (ITC), masked language modeling (MLM), and image-text matching (ITM). The authors also introduce momentum distillation to provide pseudo-labels in order to compensate for the potentially incomplete or wrong text descriptions in the noisy web training data. Using their proposed architecture and training objectives, ALBEF obtains better results than CLIP in several zero-shot and fine-tuned multimodal benchmarks, despite using orders of magnitude less images for pre-training.

Contrastive-based alignment has shown impressive zero-shot classification and retrieval results~\cite{novack2023chils, baldrati2023zero, hendriksen2022extending}, and has been successfully fine-tuned to a multitude of tasks, such as object detection~\cite{Gu2021OpenvocabularyOD}, segmentation~\cite{zhou2023zegclip} or action recognition~\cite{huang2024froster}. The same alignment objective has also been used between other modalities~\cite{cheng2020look, hager2023best, huang2023multimodal}, and with multiple modalities at the same time~\cite{girdhar2023imagebind}.

\vspace{-2mm}
\subsection{Generative Multimodal Recommendation}
\label{sec:gener-mm}

Despite their advantages, the performance of purely contrastive RS often suffers from data sparsity and uncertainty~\cite{wang2022contrastvae}. Generative models address these issues by imposing suitable structures on their latent spaces. Moreover, generative models allow for more complex recommendations, e.g., those requiring to synthesize an image. 
In what follows, we discuss thee representative generative approaches: VAEs, diffusion models, and multimodal LLMs.



\emph{Multimodal VAEs:} While VAEs (see Section~\ref{sec:vae}) could be applied directly to multimodal data, a better approach that leverages modality specific encoders and decoders trained on large corpus of data is to partition both the input and latent spaces per modality, say image and text. However, this approach reduces the multimodal VAE to two independent VAEs, one per modality.
In ContrastVAE~\cite{wang2022contrastvae}, both modalities are aligned by adding a contrastive loss between the unimodal latent representations to the ELBO objective. Experiments show that ContrastVAE improves upon purely contrastive models by adequately modeling data uncertainty and sparsity, and being robust to perturbations in the latent space.


\emph{Diffusion models}, explained in Section~\ref{sec:diffusion}, are state-of-the-art models for image generation. While they can also be used for text generation, e.g., by using a discrete latent space with categorical transition probabilities \cite{austin2021structured}, text encoders based on transformers or other sequence-to-sequence models are preferred in practice. As a consequence, multimodal models for both text and images, such as text-to-image generation models, combine text encoders with diffusion models for images. For instance, DALL-E~\cite{ramesh2022hierarchical} uses the CLIP embedding space as a starting point to generate novel images, and Stable Diffusion~\cite{rombach2022high} uses a UNet autoencoder separately pre-trained on a perceptual loss and a patch-based adversarial objective. Several works have built on and expanded diffusion models by increasing controllability of the generated results~\cite{zhang2023adding}, consistency on the generated subjects identity~\cite{ruiz2023dreambooth}, or for virtual try on~\cite{zhu2023tryondiffusion}.

\emph{Multimodal LLMs (MLLM)} provide a natural language interface for users to express their queries in multiple modalities, or even see responses in different modalities to help visualize the products. Given the complexity of training large generative models end-to-end, researchers typically assemble systems composed of discriminatively pre-trained encoders and decoders, usually connected by adaptation layers to ensure that unimodal representations are aligned. Another approach that involves little or no training is to allow a "controller" LLM to use external foundation models, or tools, to deal with the multimodal input and output~\cite{zhang2024mm}. Then, instruction tuning is an important step to make LLMs useful task solvers.
Llava~\cite{liu2024visual} is a multimodal LLM that accepts both text and image inputs, and produces useful textual responses. The authors connect a CLIP encoder with an LLM decoder using a simple linear adaptation layer.
In~\cite{liu2023improved} the authors change the connection layer from a linear projection to a two-layer MLP and obtain better results. Although MLLM research is still in its inception, some works already start using them in recommendation applications~\cite{karra2024interarec}. 

\section{Evaluating for Impact and Harm}
Evaluating RS is a complex and multifaceted task that goes beyond simply measuring a few key metrics of a single model . These systems are composed of one or more recommender models and various other ML and non-ML components, making it highly non-trivial to assess and evaluate the performance of an individual model. Moreover, these systems can have far-reaching impacts on users' experiences, opinions, and actions, which may be difficult to quantify or predict, which adds to the challenge. The introduction of Gen-RecSys further complicates the evaluation process due to the lack of well-established benchmarks and the open-ended nature of their tasks. When evaluating RS, it is crucial to distinguish between two main targets of evaluation: the system's performance and capabilities, and its potential for causing safety issues and societal harm. We review these targets, discuss evaluation metrics, and conclude with open challenges and future research directions.

\subsection{Evaluating for Offline Impact}
The typical approach to evaluating a model involves understanding its accuracy
in an offline setting, followed by live experiments.

\subsubsection{Accuracy Metrics}
The usual metrics used for discriminative tasks are recall@k, precision@k, NDCG@k, AUC, ROC, RMSE, MAE, etc. Many recent works on generative RS (e.g., \cite{kang2017visually, kang2023llms, rajput2024recommender, bao2023tallrec, he2023large}) incorporate such metrics for discriminative tasks.

For the generative tasks, we can borrow techniques from NLP. For example, the BLEU score is widely used for machine translation and can be useful for evaluating explanations\cite{geng2022recommendation}, review generation, and conversational recommendations. The ROUGE score, commonly used for evaluating machine-generated summarization, could be helpful again for explanations or review summarization. Similarly, perplexity is another metric that could be broadly useful, including during the training process to ensure that the model is learning the language modeling component appropriately \cite{lu2021revcore}.


\subsubsection{Computational Efficiency}
Evaluating computational \\efficiency is crucial for generative recommender models, both for training and inference, owing to their computational burden. This is an upcoming area of research.

\subsubsection{Benchmarks}
Many existing benchmark datasets popular in discriminative recommender models, such as Movielens~\cite{harper2015movielens}, Amazon Reviews~\cite{he2016ups}, Yelp Challenge\cite{yelp_dataset}, Last.fm~\cite{schedl2016lfm}, and Book-Crossing~\cite{ziegler2005improving}, are still useful in generative recommender models, but only narrowly. Some recent ones, like ReDial~\cite{li2018conversational} and INSPIRED~\cite{hayati-etal-2020-inspired}, are useful datasets for conversational recommendations.~\cite{deldjoo2024fairevalllm,deldjoo2024cfairllm,zhang2023chatgpt} propose benchmarks called cFairLLM and FaiRLLM, to evaluate consumer fairness in LLMs based on the sensitivity of pretrained LLMs to protected attributes in tailoring recommendations. 
We note that some benchmarks such as BigBench\cite{srivastava2023beyond} which are commonly used by the LLM community, have recommendations tasks. It will be specifically useful for the RS community to develop new benchmarks for tasks unlocked by Gen-RecSys models.

\subsection{Online and Longitudinal Evaluations}
Offline experiments may not capture an accurate picture because of the interdependence of the different models used in the system and other factors. So, A/B experiments help understand the model's performance along several axes in real-world settings. Note that \cite{wang2023recagent} proposes a new paradigm of using simulation using agents to evaluate recommender models. In addition to the short-term impact on engagement/satisfaction, the platform owners will be interested in understanding the \textit{long-term impact}. This can be measured using business metrics such as revenue and engagement (time spent, conversions). Several metrics could be used to capture the impact on users (daily/monthly active users, user sentiment, safety, harm).

\subsection{Conversational Evaluation}

BLEU and perplexity are useful for conversational evaluation but should be supplemented with task-specific metrics (e.g., recall) or objective-specific metrics (e.g., response diversity \cite{li2015diversity}). Strong LLMs can act as judges, but human evaluation remains the gold standard. Toolkits like CRSLab \cite{crslab} simplify building and evaluating conversational models, but lack of labeled data in industrial use cases poses a challenge. Some studies use LLM-powered user simulations to generate data.


\subsection{Evaluating for Societal Impact}



Previous work has investigated categories of interest for societal impacts of traditional RS \cite{milano2020recommender} and generative models \cite{weidinger2022taxonomy,bird2023typology} independently. In the context of RS literature, six categories of harms are found to be associated with RS: \textit{content}, \textit{privacy} violations and data misuse, threats to human \textit{autonomy} and well-being, \textit{transparency and accountability}, harmful \textit{social effects} such as filter bubbles, polarisation, manipulability, and \textit{fairness}. 
In addition, RS based on generative models can present new challenges  \citep{weidinger2022taxonomy,bird2023typology}:
\begin{itemize}
    \item LLMs use out-of-domain knowledge, introducing different sources of societal bias that are not easily captured by existing evaluation techniques \cite{sanner:ipm23,deldjoo2024understanding,deldjoo2024fairevalllm}.
    \item The significant computational requirements of LLMs lead to heightened environmental impacts \cite{luccioni2023power,berthelot2024environmental}. 
    \item The automation of content creation and curation may displace human workers in industries such as journalism \cite{de2023generative}, creative writing, and content moderation, leading to social and economic disruption \cite{arguedas2023automating}.
    \item Recommender systems powered by generative models may be susceptible to manipulation and could have unintended and unexpected consequences for users \citep{carroll2022estimating,kasirzadeh2023user}. 
    \item Generative recommendations can expose users to the potential pitfalls of hyper-personalization \cite{gabriel2024ethics,rillig2024ai}. 
\end{itemize}

\subsection{Holistic Evaluations}
As mentioned above, thoroughly evaluating RS for offline metrics, online performance, and harm is highly non-trivial. Moreover, different stakeholders (e.g. platform owners and users)~\cite{abdollahpouri2020multistakeholder, surer2018multistakeholder, milano2021multistakeholders} may approach evaluation differently.  The complexity of Gen-RecSys evaluation presents an opportunity for further research and specialized tools. Drawing inspiration from the HELM benchmark~\cite{liang2022holistic}, a comprehensive evaluation framework tailored for Gen-RecSys would benefit the community.

\section{Conclusions and Future Directions}
While many directions for future work have been highlighted above, the following topics constitute especially important challenges and opportunities for Gen-RecSys:

\begin{itemize}
    \item \textbf{RAG} (cf. Section 3.3), including: data fusion for multiple (potentially subjective) sources such as reviews \cite{yu2021kg, ye2023fid}, end-to-end retriever-generator training \cite{lewis2020retrieval, izacard2020leveraging, borgeaud2022improving}, and systematic studies of generative reranking alternatives \cite{qin2023large}.
    \item \textbf{Tool-augmented LLMs} for conversational recommendation, focusing on architecture design for LLM-driven control of dialogue, recommender modules, external reasoners, retrievers, and other tools \cite{friedman2023leveraging, wang2023recmind, mialon2023augmented, sparks_of_agi}, especially methods for \textit{proactive} conversational recommendation.
    \item \textbf{Personalized Content Generation} such as virtual try-on experiences \cite{zhu2023tryondiffusion}, which can allow users to visualize themselves wearing recommended clothing or accessories, improving customer satisfaction and reducing returns.
    \item \textbf{Red-teaming} -- in addition to the standard evaluations, real-world generative RS will have to undergo red-teaming (i.e., adversarial attacks) \cite{deng2023attack, wang2023aligning, shu2024attackeval} before deployment to stress test the system for prompt injections, robustness, alignment verification, and other factors.
\end{itemize}

Despite being a short survey, this work has attempted to provide a foundational understanding of the rich landscape of generative models within recommendation systems. It extends the discussion beyond LLMs to a broad spectrum of generative models, exploring their applications across user-item interactions, textual data, and multimodal contexts. It highlights key evaluation challenges, addressing performance, fairness, privacy, and societal impact, thereby establishing a new benchmark for future research in the domain.


\bibliographystyle{ACM-Reference-Format}
\bibliography{refs}

\end{document}